\documentclass[aps]{revtex4}
\usepackage{graphicx}

\begin{document}

\title{Suppression of the virtual Anderson transition in a
narrow impurity band of doped quantum well structures.}

\author{N.\,V. Agrinskaya}   \email{nina.agrins@mail.ioffe.ru}
\author{V.\,I. Kozub}
\author{D.\,S. Poloskin}
\affiliation{Ioffe Physical-Technical Institute of the Russian
Academy of Sciences, 194021, Saint Petersburg, Russia.}

\begin{abstract}

Earlier we reported an observation at low temperatures of
activation conductivity with small activation energies in strongly
doped uncompensated layers of p-GaAs/AlGaAs quantum wells. We
attributed it to Anderson delocalization of electronic states in
the vicinity of the maximum of the narrow impurity band. A
possibility of such delocalization at relatively small impurity
concentration is related to the small width of the impurity band
characterized by weak disorder. In this case the carriers were
activated from the "bandtail" while its presence was related to
weak background compensation. Here we study an effect of the
extrinsic compensation and of the impurity concentration on this
"virtual" Anderson transition. It was shown that an increase of
compensation initially does not affect the Anderson transition,
however at strong compensations the transition is suppressed due
to increase of disorder. In its turn, an increase of the dopant
concentration initially leads to a suppression of the transition
due an increase of disorder, the latter resulting from a partial
overlap of the Hubbard bands. However at larger concentration the
conductivity becomes to be metallic due to Mott transition.

\end{abstract}

\maketitle

\section{Introduction}

In our previous papers \cite{ours1}, \cite{ours2} we reported an
observation of delocalized electronic states in the vicinity of
the maximum of a narrow impurity band in GaAs/AlGaAs quantum well
structures. It is important that such a delocalization took place
at dopant concentrations significantly smaller than ones
corresponding to the Mott-Anderson criterion. Such a behavior we
described as a manifestation of the virtual Anderson transition -
because, despite of a presence of delocalized states, transport of
the majority carriers over these (occupied) states was blocked by
the Hubbard correlations. The smallness of the critical
concentration for the Anderson transition was explained by a
weakness of the random potential and, correspondingly, by a
smallness of the scatter of localized states energies. In this
case linear carrier transport was supported by minority carriers
activated from acceptors ionized by background impurities. Note
that such a behavior can be realized only 2D systems, where the
background compensating defects can be situated outside of the 2D
layer which is important for a formation of the narrow impurity
band. Since the conductivity was related to a presence of finite
(but weak) compensation and to activation of the minority carriers
situated in the bandtail to the band of delocalized states, the
corresponding activation energy was by an order of magnitude
smaller than the activation energy of the dopant. Another
mechanism of conductivity over the delocalized states was relevant
for the case of strong enough electric fields. It was related to
an impact ionization of the minority carriers to the band of
delocalized states. According to the considerations given above,
our picture corresponds to the case when the Fermi level is
situated below the mobility edge within the impurity band. One
could expect that an increase of a compensation degree, which
shifts the Fermi level towards the mobility edge, finally would
lead to purely metallic (non-activated) conductivity. However, an
increase of the compensation can lead to an increase of disorder,
i.e. to an increase of the critical concentration, and, finally,
to a suppression of the delocalization.

In its turn, as we noted earlier \cite{ours1}, an increase of the
dopant concentration can also lead to a suppression of the virtual
Anderson transition. Indeed, an increase of this concentration can
lead to an overlap of the two Hubbard bands and to appearance of
the charged centers not related to compensation. Again, the
additional charge disorder can suppress the delocalization.
However the further increase of the dopant concentration finally
leads to the Mott transition when the Fermi level reaches the
mobility edge of the upper Hubbard band. Thus, it was of interest
to study an effect of both compensating centers concentration
(introduced artificially) and of the dopant concentration on the
manifestation of the virtual Anderson transition. It is this
investigation which is reported in this paper. We will show that
an increase of compensation initially leads to a decrease of the
activation energy in agreement to the considerations given above.
The further increase of compensation leads to a complete
suppression of the virtual Anderson transition which is
accompanied by a significant broadening of the impurity band. The
latter manifests itself in an increase of the all activation
energies. Then, an increase of the dopand concentration in
uncompensated samples also initially leads to a suppression of the
virtual Anderson transition. However at large enough
concentrations the metallic state resulting from the Mott
transition is formed.

\section{Experiment}

We have studied two sets of $GaAs/Al_{0.3} Ga_{0.7} As$ quantum
well samples, grown by molecular epitaxy technique. The samples
contained 5 quantum wells with a width 15 nm separated by
$Al_{0.3} Ga_{0.7} As$ barriers with a thickness 100 nm. At the
first set of samples only quantum wells centers (5 nm) were doped
by Be with large enough concentration $(1-3)~ 10 ^{12}$ см$^{-2}$
(samples N 1,2). At the second set of samples the centers of the
quantum wells were also doped by Be, however the centers of
barriers (5 nm) were doped by compensating impurity (Si), see
Table 1 (samples 3,4). Thus, the degrees of compensation
$K=N_D/N_A<0,01$ were less than 0.01 for samples 1,2 and for
samples 3 and 4 K=0.1 and K=0.5, respectively. The acceptor
concentration in the wells was taken to be large enough to ensure
a presence of delocalized states in the center of impurity band of
$A_0$ centers (resulting from the virtual Anderson
transition)\cite{ours1}, \cite{ours2}.
\vskip0.5cm

\begin{tabular}{|c|c|c|c|c|c|c|}

\hline
  N & $N_A,cm^-2$ & $N_D,

  cm^-2$ & $p_{300K}, cm^{-2}
  $ & $\varepsilon_1$ meV  & $\varepsilon_4$,meV&
  \
 $ K=N_D/N_A$ \\
  \hline
  1-7-582 & $2,5\cdot 10^{12}$ & - & $2\cdot10^{12}$ & $-$ &- & $\leq$ 0.01\\
  2-7-580 & $1,5\cdot 10^{12}$ & , $$ & $1\cdot10^{12}$ & 21 & 2
   & $\leq$ 0.01
  \\
  3-8-291 & $1\cdot 10^{12}$ & Si, $1,5\cdot10^{11}$ & $1\cdot10^{12}$ & 21 & 2 & 0.1\\
  4-8-292 & $1\cdot10^{12}$  & Si, $5\cdot10^{11}$& $5\cdot10^{11}$ & 40 & 10  & 0.5\\

\hline
\end{tabular}

Table 1. Parameters of the studied samples

\vskip0.5cm

On Fig.1(a,b) we give the temperature curves of the conductivity
$\sigma$ and of carrier concentration P (obtained from the Hall
coefficient $R_h$) for all of the samples.

\begin{figure}[htbp]
    \centering
       \includegraphics[width=0.6\textwidth]{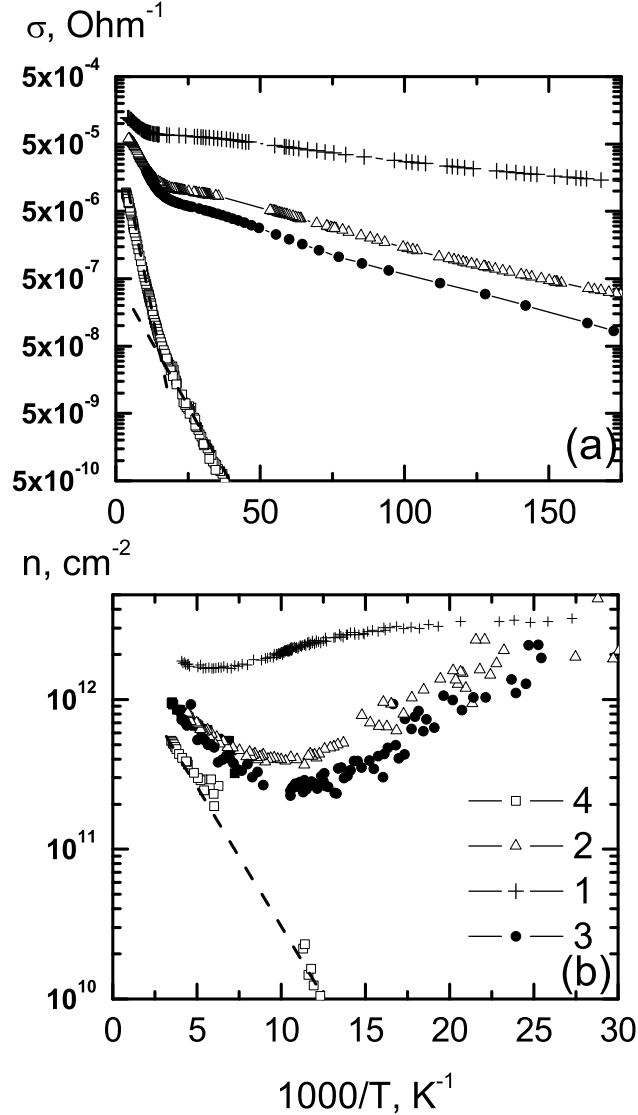}
        \caption{
        Temperature dependence of the (a) conductivity and (b)
        hole concentration for all samples listed in the table.
        \label{fig1}   }
\end{figure}

As it was noted in our previous paper \cite{ours3}, one of the
features characterizing
  the virtual Anderson transition in the impurity band is a mixed
  conductivity (over the allowed and impurity bands) which
  manifests itself as a presence of a characteristic minimum in
  temperature behavior of concentration. At higher temperatures
  there exists a part of the curve corresponding to activation of
  the carriers from the Fermi level to the allowed band - energy $\varepsilon1$
  (in more detail see about a calculation of this activation
  energy for the case of the mixed conductivity in \cite{ours3}).
  This minimum appears to be deeper with a sharp increase of the
  effective concentration and, correspondingly, with a decrease of
  the Hall mobility at temperatures 20-30 K provided the carriers
  in the allowed band and in the impurity band have opposite
  charges. When calculating the mixed concentration $Р_m$ from the
  Hall effect one should take into account that in the impurity
  band completely filled by holes the carriers can be of the
  opposite sign, their number being $P_2= N_D+P_1$. Thus the
  equation for the Hall concentration $P_H$ is modified in the
  following way:

  \begin{equation}
P_H = \frac{(N_D + P_1 + bP_2)^2}{b^2P_1 - N_D - P_1}
\end{equation}

where $P_1 = N_V \exp (\varepsilon_F/k_BT)$ is the carrier
concentration within the valence band, $b$ is a ratio of
mobilities in the allowed and impurity bands, $\mu_1$ and $\mu_2$.

Note that the r.h.s. of Eq. 1 can diverge (tend to infinity) when
the concentrations of electrons in the impurity band and
concentration of holes in the valence band are nearly equal. Thus
the Hall concentration at some temperatures can appear to be
larger, than the concentration at room temperature. The mixed
conductivity (in contrast to mixed concentration) has no maxima
(at least if one does not take into account temperature behavior
of the mobility) and is given as

\begin{equation}
\sigma_m = (P_1b + N_A - N_D - P_1)\mu_1
\end{equation}

\begin{equation}
P_H = \frac{(N_D + P_1 + bP_2)^2}{b^2P_1 - N_D - P_1}
\end{equation}

\begin{equation}
\sigma_m = (P_1b + N_A - N_D - P_1)\mu_1
\end{equation}

This region of the mixed conductivity manifests itself as a
shoulder in the conductivity temperature curves. At the lower
temperatures the temperature behavior is controlled by the
activation energy $\varepsilon_4$ (related to an activation of the
carriers from the Fermi level to the mobility edge within the
impurity band). At the weak compensations this energy is about a
halfwidth of the impurity band.

When only centers of the wells are doped, there exists the filled
impurity band of the singly occupied acceptors $A_0$, where the
compensation degree is small enough and is controlled by a
presence of  random donor impurities in the barrier ($N_D
<10^{16}$ cm$^{-3}$). The delocalized states appear in this band
at concentration $N_A \sim 10^{18}$ см$^{-3}$. In this case the
concentration of carriers within the allowed band a standard
expression for weakly compensated impurity can be used. Such a
behavior is observed for doped quantum wells without intended
compensation or at weak compensation (samples 2 and 3).
Temperature behavior of conductivity at small temperatures has an
activated character with a small energy $\varepsilon_4$. At
temperatures less than 4 K, these samples demonstrated the
breakdown behavior related to the impact ionization of the
carriers from the Fermi level to the mobility edge.

An important difference between the weakly compensated sample N2
from the uncompensated N1 is about twice less value of the
activation energy $\varepsilon_4$, as well as the breakdown
behavior at higher temperatures (around 4 K), see Fig.2 (note that
the measurements were made at constant current 1 nA).

\begin{figure}[htbp]
    \centering
       \includegraphics[width=0.6\textwidth]{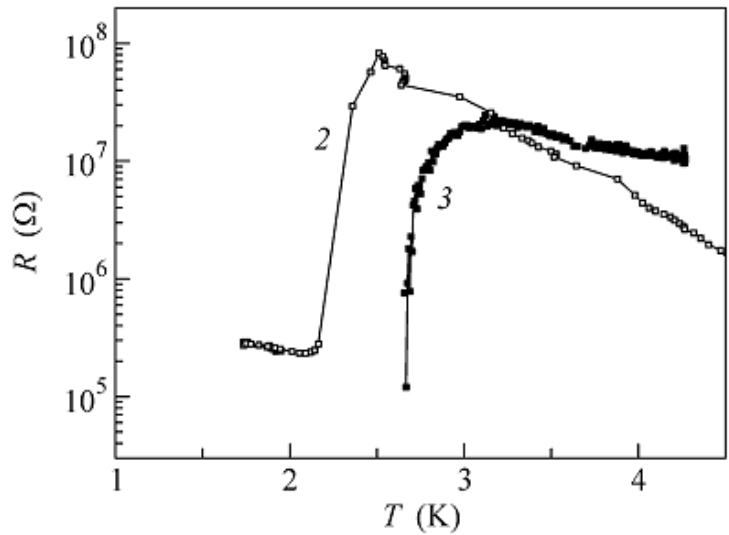}
        \caption{ Temperature dependence of the low-temperature
        conductivity of samples 2 and 3 measured in the constant
        current regime at the current of 1 na.
        \label{fig2}   }
\end{figure}

In the sample with a strong compensation degree N4 ($ N_D/N_A\sim
0,5$) the mixed conductivity as well as activated temperature
behavior at small temperatures and the breakdown behavior were not
observed. It is related to the significant increase of disorder
and, correspondingly, to the increase of the width of the impurity
band which leads to a suppression of the Anderson transition. The
activation energy $\varepsilon_1$ (obtained from the Hall data) in
this sample is $\sim$ 40 meV which exceeds significantly the
corresponding energies for weakly compensated samples. In its
turn, the low temperature conductivity behavior demonstrates
activation energy $\varepsilon_3 \sim $ 10 meV. If one relates
this energy to the nearest neighbor hopping, it corresponds to the
halfwidth of the impurity band. An increase of the energy
$\varepsilon_1$ for the strong compensation limit, according to
\cite{ES}, is equal to $\varepsilon_1=\varepsilon_0+
\varepsilon_3$, which for $\varepsilon_0$ = 30 meV gives the
observed energy $\varepsilon_1$ = 40 meV. The variable range
hopping was not observed due to very large values of resistance at
low temperatures.

With an increase of the doping level the sample N1 demonstrates
weaker temperature dependencies of the conductivity and Hall
effect (Fig.1). This sample exhibits effects of weak localization
in temperature behavior of conductivity as well as in
magnetoresistance (MR) (Fig.3). At low temperatures in weak
magnetic fields MR appears to be positive (antilocalization), then
a transition to negative MR is observed (weak localization). This
behavior correlates to the one of quasimetallic samples (see
\cite{agrinMR}). Such a behavior evidences the Mott transition
(overlapping of the upper and lower Hubbard bands).

\section{Discussion}

1. Role of compensation.

As it was noted earlier, one could expect that an increase of
compensation degree, shifting the Fermi level $\mu$ within the
impurity band towards the band of delocalized states could finally
to nullify the activation energy $\varepsilon_4$ which would mean
a transition to metallic state. Indeed, assuming that the density
of states of the impurity band has Gaussian shape, for small
concentrations of the compensating donors one obtains
\begin{equation}
\int_{-\infty}^{\mu} g_0 \exp -(\frac{\varepsilon_0 -
\varepsilon'}{\Delta \varepsilon_0})^2 {\rm d} \varepsilon' = N_d
\end{equation}
where $\varepsilon_0$ corresponds to a center of the impurity
band, $\Delta \varepsilon_0$ is the halfwidth of the impurity band
at the limit of small $N_d$. The latter is controlled mostly by
deformational effects of non-Coulombic nature and by the order of
magnitude coincides with energy $\varepsilon_4$ observed for
weakly compensated samples. In thius case the shift of the Fermi
level is given as
\begin{equation}
\frac{\partial \mu}{\partial n_d} = g_0^{-1} \exp
(\frac{\varepsilon_0 - \mu}{\Delta \varepsilon_0})^2
\end{equation}
It is seen that for small $N_d$ when the Fermi level is deep
within the tail of the impurity band, an increase of $N_d$
initially leads to strong shift of $\mu$ and, correspondingly, to
a decrease of $\varepsilon_4$. However an increase of the Coulomb
disorder leads to the broadening of the impurity band which can be
estimated as
\begin{equation}\label{dis}
\Delta \varepsilon_C \sim \frac{e^2N_d^{1/2}}{\kappa}
\end{equation}
Correspondingly, finally this broadening can exceed the value of
$\Delta \varepsilon_0$. According to considerations, given in our
papers \cite{ours1}, \cite{ours2}, the criterion for Anderson
transition for significantly compensated samples obtains a form
\begin{equation}\label{Mott1}
N_A a^2 \geq \frac{\alpha}{\ln(\varepsilon_0/(\Delta
\varepsilon_C))}
\end{equation}
where $N_A$ is a concentration of dopant acceptors, $a$ is the
localization length while $\alpha$ is of the order of unity.

This equation can be compared with a criterion of Anderson
transition in non-compensated samples :
\begin{equation}\label{Mott2}
N_A a^2 \geq \frac{\alpha}{\ln(\varepsilon_0/(\Delta
\varepsilon_0))}
\end{equation}
Since the mobility edge is expected to be situated in the vicinity
of the band center, an increase of the compensation can shift the
sample to the metallic state only at small degrees of
compensation. When the Fermi level reaches a vicinity of the band
center it inevitably means an icrease of $\Delta \varepsilon_C$ up
to the values significantly exceeding $\varepsilon_0$. As it is
seen from the comparison of the criteria \ref{Mott1}and
\ref{Mott2}, this leads (for fixed concentration $N_A$) to a
suppression of the virtual Anderson transition.

As it is seen from the estimate \ref{dis}, at compensating donor
concentrations $\sim 10 ^{-11}$ cm$^{-2}$  the magnitude of the
disorder potential is of the order of $\sim 6 meV$. Thus in the
sample 3 with a small degree of compensation with the
concentration of the compensating donors given above the energy
$\Delta \varepsilon_C$  appears to be comparable to the energy
$\Delta \varepsilon_0$ (characterizing the bandwidth without
compensation). Indeed, according to our estimates (\cite{ours1},
\cite{ours2}) the energy $\Delta \varepsilon_0$ is of the order of
$6 meV$. This estimates is compatible to the activation energy
$\varepsilon_4 \sim 2 meV$.

This scenario completely agrees with experiment. In particular,
for moderate degree of compensation ($K = 0.1$, sample 3) the
criterion of the Anderson transition \ref{Mott1} is still met, and
the observed picture is similar to the one of the sample 1. Note
that despite of the expected shift of the chemical potential
towards the center of the band, it does not lead to a significant
decrease of the activation energy. This is due to the fact that
the effect of the shift of $\mu$ is compensated by the effect of
the broadening of the very impurity band, as well as by a
narrowing of the band of Anderson-delocalized states due to an
increase of disorder. However one can not exclude a possibility
that at some relation between the dopant concentration and the
degree of compensation the real Anderson transition can occur in
the impurity band. Namely, the real quasimetallic conductivity can
be achieved when the Fermi level reaches the mobility edge.
Indeed, one of our samples demonstrated such a behavior - at the
dopant concentration $5 \cdot 10^{11}$ cm$^{-2}$ and weak
compensation the conductivity and magnetoresistance demonstrated
weak localization behavior \cite{agrinMR}.

At the same time for strongly compensated sample 4 the activation
energy $\varepsilon_1$ (obtained from the Hall measurements) is
$\sim$ 40 meV which significantly exceeds the corresponding
energies for weakly compensated samples. Then, at low temperatures
the temperature behavior of conductivity exhibits activation
energy $\varepsilon_3 \sim $ 10 meV. If one attributes this energy
to the nearest neighbor hopping, it also coincides with a halwidth
of the impurity band.

\begin{equation}\label{Mott2}
N_A a^2 \geq \frac{\alpha}{\ln(\varepsilon_0/(\Delta
\varepsilon_0))}
\end{equation}

Thus an absence of the virtual Anderson transition as well as a
significant increase of the energies $\varepsilon_1$ and
$\varepsilon_3$ at strong compensation is explained by a strong
increase of the bandwidth due to the Coulomb effects. In this case
$\varepsilon_1=\varepsilon_0+ \varepsilon_3$, which for
$\varepsilon_0$ =30 meV gives the observed energy
$\varepsilon_1$=40 meV.

2. Role of the dopant concentration.

It is of some paradox that an increase of the dopant concentration
(for a given small degree of compensation) should initially lead
to a suppression of the virtual Anderson transition. Indeed, the
increase of concentration leads to a broadening of the impurity
band due to overlapping of the wavefunctions of the neighboring
cites which we denote as $\varepsilon_T$. If $\Delta \varepsilon_T
> \Delta \varepsilon_0$ one expects an overlap of the tails of the
two Hubbard bands. This fact leads to an appearance of the charged
states within the impurity band and, correspondingly, of the
related disorder potential. The latter can be estimated by a
standard expression \ref{dis}, however instead of the
concentration of the charged centers one should insert the
concentration of doubly occupied $N_A^+$ centers resulting from
the overlap of the Hubbard bands.  Thus, according to \ref{Mott1}
it initially leads to a suppression of delocalization. However one
has in mind that the overlap of the Hubbard bands exponentially
depends on the dopant concentration while the overlap mentioned
above finally leads to the Mott transition and, correspondingly,
to non-activated metallic conductivity. It is such a behavior
which is observed for strongly doped sample 1. This sample
demonstrates weak temperature behavior of conductivity and Hall
effect while low-temperature MR demonstrates effects of weak
localization and antilocalization (at weak magnetic fields),
Fig.3.

\begin{figure}[htbp]
    \centering
       \includegraphics[width=0.6\textwidth]{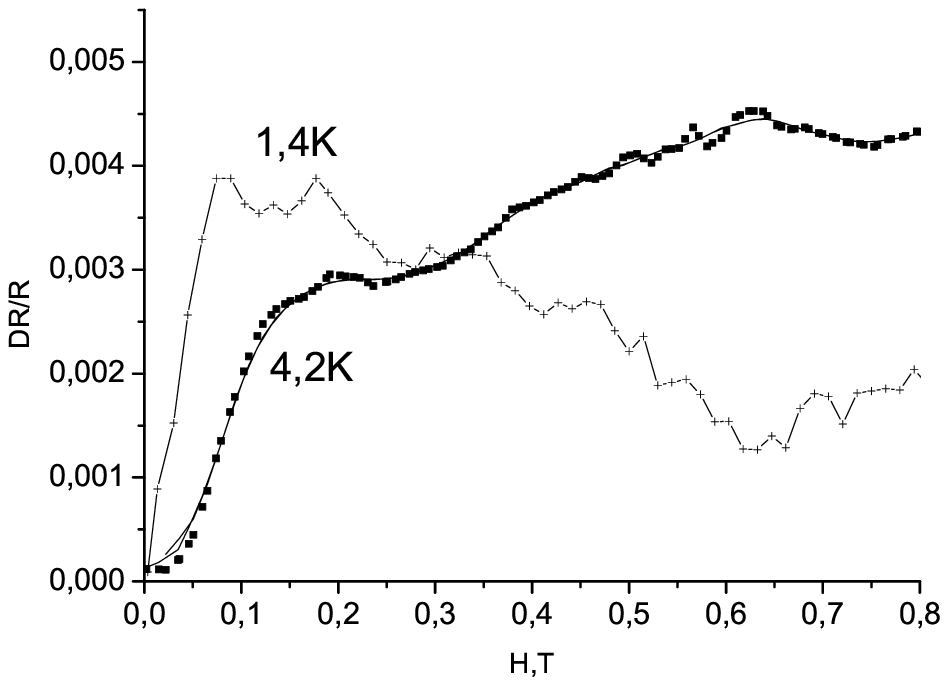}
        \caption{ Magnetoresistance curves for sample 1
        \label{fig.3}   }
\end{figure}

This correlates with a behavior of dirty metals. As it was noted
above. this behavior results from the Mott transition (that is
from the strong overlap of the upper and lower Hubbard bands).

\section{Conclusion}

Thus we have shown that an increase of the compensation degree in
the samples, demonstrating virtual Anderson transition, initially
leads to an increase of the low temperature conductivity. However
a further increase of the degree of compensation leads to a
suppression of the transition (and, correspondingly, to a decrease
of the conductivity) due to an increase of the disorder potential.
In its turn, an increase of the dopant concentration initially can
also lead to a suppression of the Anderson transition, however at
large dopant concentration the Mott transition takes place, and
the sample becomes to be metallic. Since we predicted earlier such
a behavior for the virtual Anderson transition \cite{ours1},
\cite{ours2}, the results reported in this paper give an
additional support to our model.

\section{
Acknowledgements} The paper was supported by RFBR foundation
(Project
10-02-00544).

\end{document}